\documentclass[showpacs,pre]{revtex4}

\usepackage{graphicx}
\usepackage{amsmath}
\pdfoutput=1

\begin{document}

\title{The effect of Coulombic friction on spatial displacement statistics}

\date{\today}

\author{Andreas M.~Menzel}
\altaffiliation[Current address: ]{Max Planck Institute for Polymer Research, P.O.~Box 3148, 55021 Mainz, Germany}
\email[email: ]{menzel@mpip-mainz.mpg.de}
\affiliation{Department of Physics, University of Illinois at Urbana-Champaign, Loomis Laboratory of Physics, 1110 West Green Street, Urbana, IL 61801, USA}
\author{Nigel Goldenfeld}
\affiliation{Department of Physics, University of Illinois at Urbana-Champaign, Loomis Laboratory of Physics, 1110 West Green Street, Urbana, IL 61801, USA}

\begin{abstract}
The phenomenon of Coulombic friction enters the stochastic description of dry friction between two solids and the statistic characterization of vibrating granular media. Here we analyze the corresponding Fokker-Planck equation including both velocity and spatial components, exhibiting a formal connection to a quantum mechanical harmonic oscillator in the presence of a delta potential. Numerical solutions for the resulting spatial displacement statistics show a crossover from exponential to Gaussian displacement statistics. We identify a transient intermediate regime that exhibits multiscaling properties arising from the contribution of Coulombic friction. The possible role of these effects during observations in diffusion experiments is shortly discussed.
\end{abstract}

\pacs{05.40.-a, 05.10.Gg, 62.20.Qp, 68.35.Fx}

% Stochastic models in statistical physics and nonlinear dynamics, 05.10.Gg
% Stochastic analysis, 02.50.Fz
% Stochastic processes, 05.40.-a
% Friction mechanical properties of solids, 62.20.Qp
% Fokker-Planck equation statistical physics, 05.10.Gg
% Langevin method, 05.10.Gg
% Diffusion in liquids, 66.10.C-
% Diffusion at solid surfaces and interfaces, 68.35.Fx
% Harmonic oscillators, 03.65.Ge
% Scaling phenomena in complex systems, 89.75.Da

\maketitle

\section{Introduction}

Since Langevin's early investigations \cite{langevin1908theory,lemons1997paul}, the motion of mesoscopic particles has been studied by equations basically of the Newtonian type, supplemented by random stochastic force terms. Through well-known and straightforward formalisms, these ``Langevin equations'' are connected to the continuum descriptions of the Fokker-Planck type \cite{fokker1914mittlere,planck1917uber,zwanzig2001nonequilibrium}. The latter give access to the corresponding stochastic distribution functions, stationary or time-dependent.

In the simplest set-up within this framework, the velocity component of a one-dimensional stochastic motion is investigated \cite{rayleigh1891}. Here, we consider the second-simplest example. That is, we include both the velocity and spatial component of the one-dimensional motion of a single particle \cite{zwanzig2001nonequilibrium}. Numerous studies are related to this scenario, when the particles are additionally exposed to nonharmonic spatial potentials \cite{ambegaokar1969voltage,dieterich1977diffusion,risken1978correlation,praestgaard1981model}. On the contrary, however, investigations on particles exposed to forces that are nonlinear in the velocity component are much less frequently encountered \cite{kramers1940brownian}. An example of the latter type forms the subject of this paper.

More precisely, we refer to frictional forces of the Coulombic type \cite{persson2000sliding}. They form a very basic example of nonlinear frictional behavior, in contrast to conventional linear viscous frictional forces. Only very recently have they been included into the stochastic characterization of the Langevin and Fokker-Planck type by de Gennes \cite{gennes2005brownian}, who studied the dry friction between two solids, and Kawarada and Hayakawa \cite{kawarada2004non}, who were interested in the statistics of vibrating granular media. In both studies, the authors consider only the velocity component. The same is true for a more formal study using the path integral formalism \cite{baule2010path}.

Examples of studies in which the impact of Coulombic frictional forces on the spatial displacement statistics is taken into account are even rarer. So far, such investigations have been performed numerically in the context of contact line motion of water droplets on vibrating solid substrates \cite{mettu2010stochastic} and of stick-slip motions of solid particles on vibrating substrates \cite{goohpattader2010diffusive}.

The purpose of this paper is to present results of further analytical considerations of the underlying equations. We find that the velocity-dependent part of the corresponding Fokker-Planck equation is formally connected to the Schr\"odinger equation for the quantum mechanical harmonic oscillator in the presence of a delta potential. The spatial distribution function is then obtained by numerically solving the corresponding Brinkman hierarchy \cite{brinkman1956brownian}. We will present the details of this analysis in section \ref{analytical} of this paper, after a short review of the underlying equations in section \ref{equations}.

Starting from a sharp spatial distribution, a direct numerical integration of the Fokker-Planck equation in time reveals a crossover from a subsequent exponential to a Gaussian spatial distribution function, as shown in section \ref{numerical}. The exponential tails in the spatial distribution function result from the influence of the Coulombic friction. This has been found before, using a less direct way of numerical calculation \cite{mettu2010stochastic,goohpattader2010diffusive} compared to the one applied here. We show that a data collapse of the resulting spatial distribution curves at different times is not possible by a simple rescaling procedure. Taking into account higher moments of the spatial distribution function, we identify an intermediate regime of multiscaling.

Originally, the concept of Coulombic friction was introduced to describe the interactions between rigid solids. However, as pointed out in section \ref{discussion}, Coulombic friction may also play a role in systems that feature an apparently regular diffusive behavior. This is because the mean square displacement still increases linearly in time. Higher order moments of the spatial displacement distribution function must be analyzed to identify the impact of Coulombic friction. 

The last section is left for the conclusions.

\section{\label{equations}Stochastic equations}

Coulombic friction was introduced into the Langevin equation by considering the term $-\Delta\sigma(v)$ in addition to the viscous frictional force \cite{gennes2005brownian}. Here, $\sigma(v)$ is the sign-function
\[
        \sigma(v)=\left\{
            \begin{array}{cc} +1 & \qquad i\!f \quad v>0, \\
                              0  & \qquad i\!f \quad v=0, \\
                              -1 & \qquad i\!f \quad v<0,
            \end{array}
          \right.
\]
and $\Delta$ is the strength of the Coulombic frictional force (not the Laplace operator). Including the spatial component, we obtain for the one-dimensional stochastic motion of a single particle the coupled system of equations
\begin{eqnarray}
    m\,\frac{dv}{dt} & = & {}-m\frac{v}{\tau}-\Delta\sigma(v)+\gamma(t),
    \label{langevinv} \\
    \frac{dx}{dt}  & = & v.  \label{langevinx}
\end{eqnarray}
$m$ is the mass of the particle, and the first term on the right hand side of eq.~(\ref{langevinv}) includes the viscous force with $\tau$ the corresponding relaxation time. The last term, $\gamma(t)$, gives the stochastic force, which is assumed to be $\delta$-correlated and of Gaussian type: $\langle\gamma(t)\rangle=0$, $\langle\gamma(t)\,\gamma(t')\rangle = 2Kk_BT\delta(t-t')$, $K$ characterizing the strength of the force, $k_B$ being the Boltzmann constant, and $T$ the temperature. We can scale out $m$, $\tau$, and $k_BT$ through the transformation $v'=(m/k_BT)^{1/2}\,v$, $x'=1/\tau\,(m/k_BT)^{1/2}\,x$, $t'=t/\tau$, $\Delta'=\tau/(mk_BT)^{1/2}\,\Delta$, and $K'=\tau/(mk_BT)\,K$, where primes are neglected in the following.

\begin{figure}
\includegraphics[width=8.5cm]{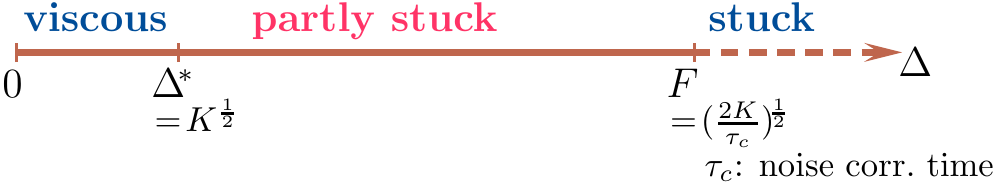}
\caption{\label{fig_phasediagram}(Color online) Qualitative ``line phase diagram'' following de Gennes \cite{gennes2005brownian}. The three different regimes ``viscous'', ``partly stuck'', and ``stuck'' are shown as a function of the strength of the Coulombic frictional force $\Delta$.}
\end{figure}

Comparing the magnitude of the terms on the right hand side of eq.~(\ref{langevinv}), de Gennes identified three different regimes \cite{gennes2005brownian}. We summarize them in the ``line phase diagram'' in fig.~\ref{fig_phasediagram}.
If $\Delta$ is small enough, the viscous frictional forces dominate and the behavior of the system is similar to the case originally studied by Langevin \cite{langevin1908theory,lemons1997paul}. Above a crossover value $\Delta^*=K^{1/2}$, the Coulombic dry friction dominates the viscous force. This regime is called the ``partly stuck regime'', if the particle is not yet completely stuck. The latter happens beyond a value $F=(2K/\tau_c)^{1/2}$, where Coulombic friction also outweighs the stochastic force $\gamma(t)$. Here, $\tau_c$ is the correlation time of the stochastic noise. In a strict sense, this regime is never reached when the spectrum of the stochastic force is purely white ($\tau_c\rightarrow0$).

Numerically integrating the coupled Langevin equations (\ref{langevinv}) and (\ref{langevinx}) forward in time requires a specific update scheme. In principle, the time steps must be infinitely small in order to respect the singular behavior of the $-\Delta\sigma(v)$ term at $v=0$. We will concentrate on the continuum picture in the following by investigating the corresponding Fokker-Planck equation.

\begin{figure}
\includegraphics[width=8.5cm]{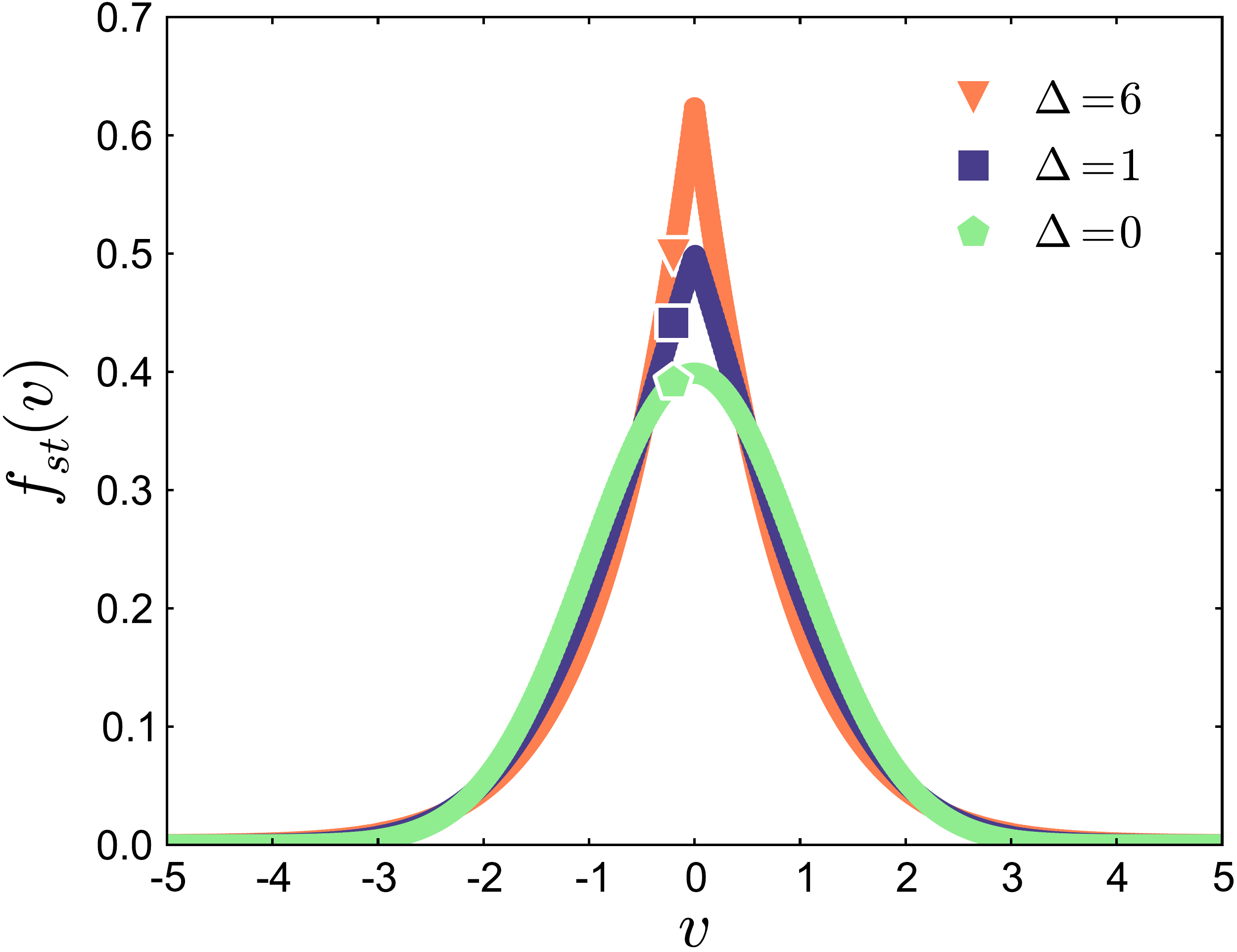}
\caption{\label{fig_fst}(Color online) Examples for the shape of the velocity-dependent stationary solution of the Fokker-Planck equation (\ref{FP}). With increasing strength $\Delta$ of the Coulombic frictional force, the cusp singularity at $v=0$ becomes more and more pronounced.}
\end{figure}

Scaling out $m$, $\tau$, and $k_BT$, this equation becomes
\begin{equation}
    \partial_t f = \left\{-v\partial_x  +\partial_v\left(v+\Delta\sigma(v)\right) + K\partial_v^2\right\} f.  \label{FP}
\end{equation}
In our case, $f=f(x,v,t)$ is the space-, velocity-, and time-dependent probability distribution function. The stationary, velocity-dependent solution of this equation reads
\begin{equation}\label{fst}
    f_{st}(v)=\frac{e^{-\frac{1}{K}\left(\frac{v^2}{2}+\Delta|v|\right)}}{\sqrt{2\pi K}
      e^{\frac{\Delta^2}{2K}}\left(1-\text{erf}\left\{\frac{\Delta}{\sqrt{2K}}\right\}\right)}.
\end{equation}
As expected, we retrieve the Gaussian shape in the absence of the Coulombic frictional contribution, $\Delta=0$. Increasing the Coulombic frictional parameter $\Delta$ leads to the emergence of a cusp singularity at $v=0$ (see fig.~\ref{fig_fst}).

\begin{figure}
\includegraphics[width=8.5cm]{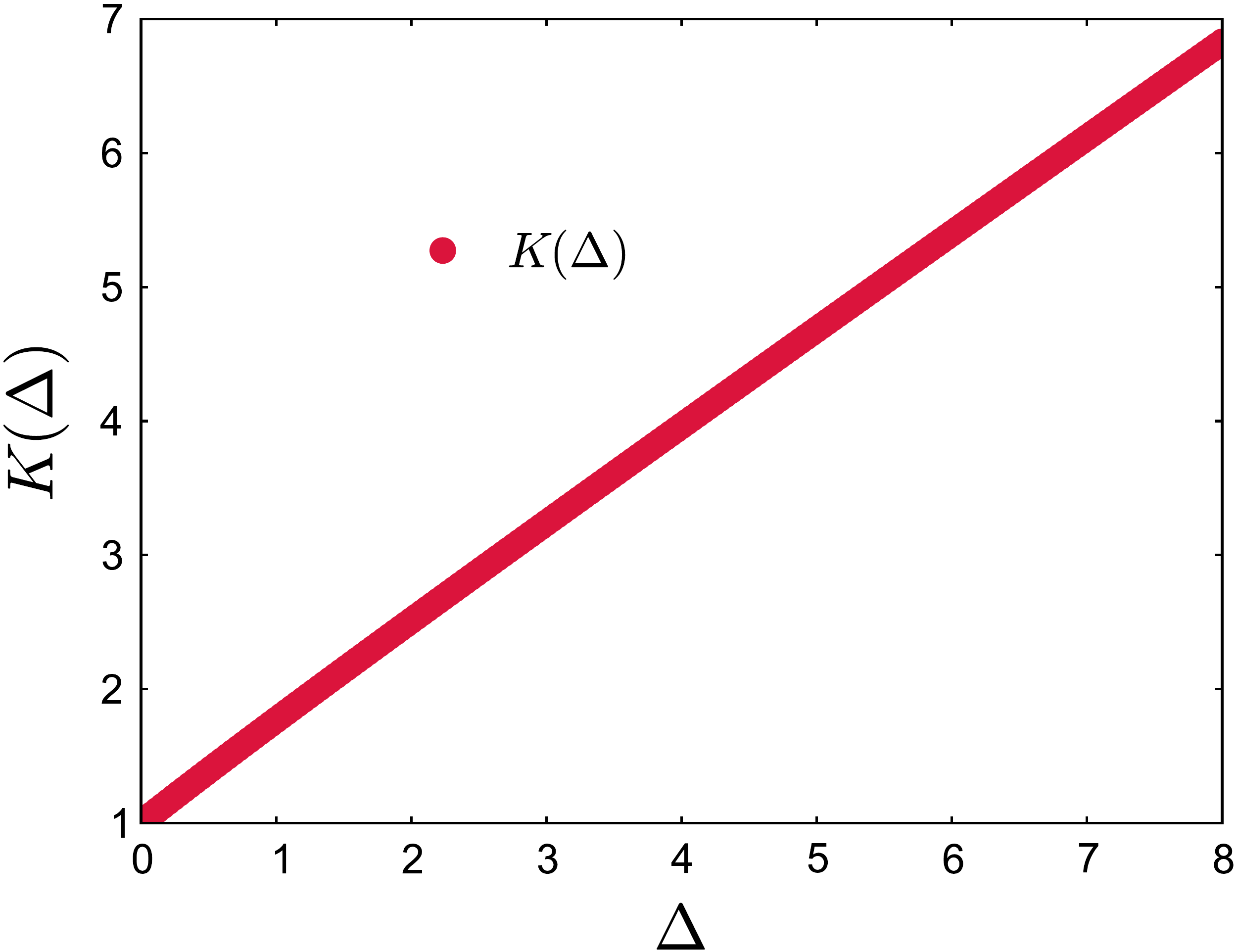}
\caption{\label{fig_flucdiss}(Color online) Numerical solution of eq.~(\ref{flucdiss}): strength of the stochastic force $K$ as a function of the strength of the Coulombic frictional force $\Delta$.}
\end{figure}

We will consider two different scenarios to fix the value of the strength of the stochastic force, $K$. On the one hand, we may define a particle temperature in the stationary state from the averaged square velocity $\langle v^2\rangle$. $f_{st}(v)$ from eq.~(\ref{fst}) is used to calculate this average. If we set the resulting particle temperature equal to the overall temperature of the system, we obtain a fluctuation dissipation relation between the strength of the stochastic force, $K$, and the strength of the Coulombic frictional force, $\Delta$:
\begin{equation}
    K+\Delta^2-\frac{\sqrt{\frac{2K}{\pi}}\Delta e^{-\frac{\Delta^2}{2K}}}{1-\text{erf}\left\{\frac{\Delta}{\sqrt{2K}}\right\}}=1.  \label{flucdiss}
\end{equation}
Numerical solution of this equation reveals a roughly linear relation between $\Delta$ and $K$ in the interesting parameter regime, as shown in fig.~\ref{fig_flucdiss}. 
Further remarks on this relation are included in the appendix. 

On the other hand, we may set $K=1$. This is the value corresponding to the fluctuation dissipation theorem for conventional Brownian motion in the absence of Coulombic friction. The latter choice implies that the dissipative process of Coulombic friction does not alter the nature of the stochastic force on the particle.

We will come back to these two different scenarios in section \ref{discussion}.
In both cases, we are left with only one independent parameter $\Delta$. We remark that the parameter $K$ can also be scaled out of eq.~(\ref{FP}) as indicated at the beginning of the next section.

\section{\label{analytical}Analytical considerations}

We now turn to the further investigation of eq.~(\ref{FP}) pointing out analytical relations not recognized so far. For this purpose, we note that the parameter $K$ can be scaled out through the transformation $\hat{x}=x/\sqrt{K}$, $\hat{v}=v/\sqrt{K}$, and $\hat{\Delta}=\Delta/\sqrt{K}$. This implies $\hat{f}(\hat{x},\hat{v},t)=K f(x,v,t)$ due to normalization. In favor of readability the $\hat{}$ will be omitted.

The Fokker-Planck operator on the right hand side of eq.~(\ref{FP}) can then be split into a reversible and a non-reversible part,
\begin{eqnarray}
  L_{rev} & = & -v\partial_x, \\
  L_{ir}  & = & \partial_v\!\left(v+\Delta\sigma(v)\right) + \partial_v^2.
\end{eqnarray}
Via the usual transformation \cite{risken1996fokker}, $L\rightarrow\bar{L}=\sqrt{f_{st}}^{-1}L\sqrt{f_{st}}$, we obtain
\begin{eqnarray}
  \bar{L}_{rev} & = & -{v}\partial_{{x}}, \\
  \bar{L}_{ir}  & = & \frac{1}{2}\left[1+2{\Delta}\delta({v})\right]-\frac{1}{4}\left[{v}+{\Delta}\sigma({v})\right]^2+\partial_{{v}}^2.   \label{barLir}
\end{eqnarray}
This result is based on the relation $\partial_v\sigma(v)=2\delta(v)$. Formally, the latter follows from $\sigma(v)=2\Theta(v)-1$, where the Heaviside function $\Theta(v)$ is given by $\Theta(v>0)=1$, $\Theta(v<0)=0$, and $\Theta(v=0)=\frac{1}{2}$.

Now, $\bar{L}_{ir}$ in eq.~(\ref{barLir}) is Hermitian. This makes it possible for us to find the solutions to the eigenvalue problem
\begin{equation}
  \bar{L}_{ir}\psi_{\mu}(v)=-\mu\,\psi_{\mu}(v),  \label{ewgl}
\end{equation}
and then expand the transformed probability distribution function $\bar{f}(x,v,t)$ as
\begin{equation}
  \bar{f}(x,v,t)=\sum_{\mu}c_{\mu}(x,t)\psi_{\mu}(v). \label{expansion}
\end{equation}
The advantage of this formulation is that the variables $(x,t)$ can be separated from the velocity $v$. Inserting the expansion (\ref{expansion}) into the transformed version of eq.~(\ref{FP}) leads to a hierarchy of coupled partial differential equations for the expansion coefficients $c_{\mu}(x,t)$ \cite{risken1996fokker,brinkman1956brownian}. The benefit of this procedure results from noting that $c_0(x,t)$ corresponds to the time-dependent spatial distribution function \cite{risken1996fokker},
\begin{equation}\label{c0}
  c_0(x,t)= \int_{-\infty}^{\infty} f(x,v,t)\,dv,
\end{equation}
the quantity we are looking for.

We derive the solutions to the eigenvalue problem eq.~(\ref{ewgl}) by first concentrating on the regime $v\geq 0$. A simple transformation $\tilde{v}:=v+\Delta$ ($\tilde{v}\geq\Delta$) leads to
\begin{equation}
  {}-\partial_{\tilde{v}}^2\psi_{\mu}(\tilde{v})+\left\{\frac{1}{4}\tilde{v}^2-\Delta \,\delta(\tilde{v}-\Delta)\right\}\psi_{\mu}(\tilde{v})={}\left(\mu+\frac{1}{2}\right)\,\psi_{\mu}(\tilde{v}).
\end{equation}
This transformation shows that the Hamiltonian $\bar{L}_{ir}$ is related to the case of a quantum mechanical harmonic oscillator in the presence of an additional pinning $\delta$-potential.

The harmonic oscillator potential and the pinning $\delta$-potential separately found their way into virtually every introductory course on quantum mechanics. It is amusing to note that the combination of both cases appears only relatively recently in the literature \cite{avakian1987spectroscopy,janke1988statistical,patil2006harmonic}. Performing an analysis similar to the one presented in ref.~\cite{janke1988statistical} ($z_0=\Delta$ and $-a=\Delta$ in ref.~\cite{janke1988statistical}), and including the regime $v\leq0$, we find: (1) Even eigenfunctions are given by $\psi_{\mu}(v)=C\,D_{\mu}(|v|+\Delta)$, the corresponding eigenvalues $\mu$ are determined by the relation $D_{\mu+1}(\Delta)=0$. (2) Uneven eigenfunctions are given by $\psi_{\mu}(v)=C\,\sigma(v)\,D_{\mu}(|v|+\Delta)$, the corresponding eigenvalues $\mu$ are determined by the relation $D_{\mu}(\Delta)=0$. (3) The smallest eigenvalue is given by $\mu_0=0$, with the eigenfunction $\psi_0(v)=[f_{st}(v)]^{1/2}$.
Here, $D_{\mu}(v)$ are the parabolic cylindrical functions, and $C=[\int_{-\infty}^{\infty}D_{\mu}(|v|+\Delta)^2\,dv]^{-1/2}$. The eigenfunctions $\psi_{\mu}(v)$ form an orthonormal set. In the limiting case of $\Delta\rightarrow 0$ they correctly tend to the eigenfunctions of the harmonic oscillator denoted by $\phi_n(v)$ in the following.

Using these results and expansion (\ref{expansion}) in the transformed eq.~(\ref{FP}), we obtain
\begin{equation}
  \sum_{\mu}\,\psi_{\mu}(v)\,\partial_t c_{\mu}(x,t)=-\sum_{\mu}\,\mu\,\psi_{\mu}(v)\,c_{\mu}(x,t) -\sum_{\mu}\,v\,\psi_{\mu}(v)\,\partial_x c_{\mu}(x,t).
\end{equation}
The separation of the variable $v$ from $(x,t)$ is not completed due to the factor $v$ in the last term.

For $\Delta=0$, this case was solved by noting that \cite{risken1996fokker}
\begin{equation}
  \bar{L}_{rev}\,\bar{f}=-v\partial_x\,\bar{f}=-[b^+\partial_x+b\,\partial_x]\,\bar{f}.
\end{equation}
$b^+$ and $b$ denote the creation and annihilation operators corresponding to the case of the harmonic oscillator. By climbing in the harmonic oscillator spectrum, the factor $v$ can be suppressed since
\begin{eqnarray}
  b^+ \phi_n(v) &=& \left[-\partial_v+\frac{1}{2}v\right]\,\phi_n(v)=\sqrt{n+1}\,\phi_{n+1}(v), \\
  b \,\phi_n(v) &=& \;\left[\partial_v+\frac{1}{2}v\right]\,\phi_n(v)\;=\sqrt{n}\,\phi_{n-1}(v).
\end{eqnarray}

When $\Delta\neq0$, an analogous treatment is not possible, since the eigenvalues are not separated by integer values. We therefore expand the eigenfunctions $\psi_{\mu}(v)$ into the harmonic oscillator eigenfunctions $\phi_n(v)$, then apply the creation and annihilation operators to get rid of the factor $v$, and finally expand the $\phi_n(v)$ back into our eigenfunctions $\psi_{\mu}(v)$. This leads to a Brinkman hierarchy of the form
\begin{equation}
  \partial_tc_{\mu}(x,t)={}-\mu\,c_{\mu}(x,t)-\sum_{\nu}\,e_{\mu\nu}\,\partial_xc_{\nu}(x,t)
\end{equation}
with the expansion coefficients
\begin{equation}\label{emunu}
  e_{\mu\nu}=\sum_m\,\langle\phi_m|\psi_{\nu}\rangle\,\left\{\sqrt{m}\,\langle\psi_{\mu}|\phi_{m-1}\rangle\,+\,\sqrt{m+1}\,\langle\psi_{\mu}|\phi_{m+1}\rangle\right\}.
\end{equation}
For $\Delta\rightarrow0$, these expansion coefficients correctly tend to the analytical harmonic oscillator solutions, $e_{\mu\nu}\rightarrow\sqrt{\nu}\,\delta_{\mu,\nu-1}+\sqrt{\nu+1}\,\delta_{\mu,\nu+1}$, with $\delta_{\alpha\beta}$ the Kronecker delta. $e_{\mu\nu}=0$ if both, $\mu$ and $\nu$, are either even or uneven. The latter excludes advective terms of the same order.

\begin{figure}
\includegraphics[width=8.5cm]{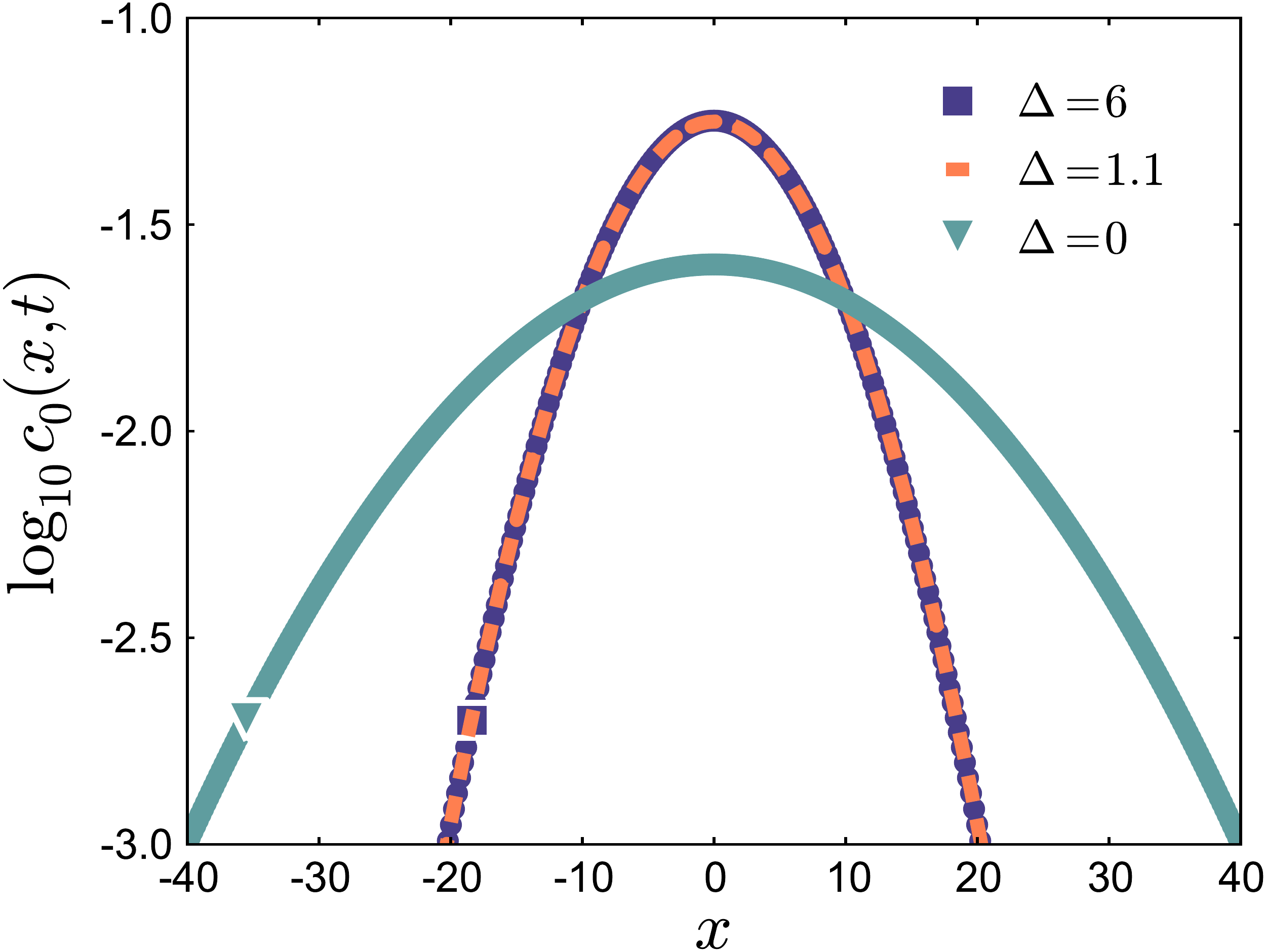}
\caption{\label{fig_c0}(Color online) Semilogarithmic plot of the spatial distribution functions $c_0(x,t)$, calculated numerically from the corresponding Brinkman cascades. The narrower (blue--solid and orange--dashed) curves correspond to two different cases of Coulombic friction ($\Delta=6$, $K\approx 5.40$ and $\Delta=1.1$, $K=1$, respectively). No Coulombic friction was present in the case of the broader (grayish) curve ($\Delta=0$, $K=1$), which therefore represents the case of conventional Brownian motion. All three curves follow from the same narrow Gaussian initial distribution after an equal amount of iterating time steps. (Technical details: variance of the initial Gaussian spatial distribution $\sigma^2=0.05$; 5000 lattice points of distance $dx=0.1$; 25000 time steps of step size $dt=0.005$. In the expansion (\ref{emunu}) the first $40$ harmonic oscillator eigenfunctions were used to calculate the corresponding coefficients $e_{\mu\nu}$. The shape of the curves is obtained after initial transient exponential tails have moved outward.)}
\end{figure}

As an example, we numerically solved the Brinkman hierarchy up to third order for two different cases of Coulombic friction. First, we set $\Delta=6$. The value of the strength $K$ of the stochastic force was chosen as $K\approx5.40$ according to the fluctuation dissipation relation (\ref{flucdiss}). Second, we set $\Delta=1.1$. Here, we chose $K=1$ corresponding to its value for a freely moving Brownian particle. We started from a narrow Gaussian spatial distribution and then iterated it forward in time to find the respective spatial distribution function $c_0(x,t)$ (see eq.~(\ref{c0})). In fig.~\ref{fig_c0}, we compare to the case of conventional diffusion without Coulombic friction ($\Delta=0$, $K=1$) but with identical initial conditions. The effect of the Coulombic frictional force term is obvious, and the two chosen cases of Coulombic friction lead to similar results.

\section{\label{numerical}Direct numerical integration}

\begin{figure}
\includegraphics[width=17.cm]{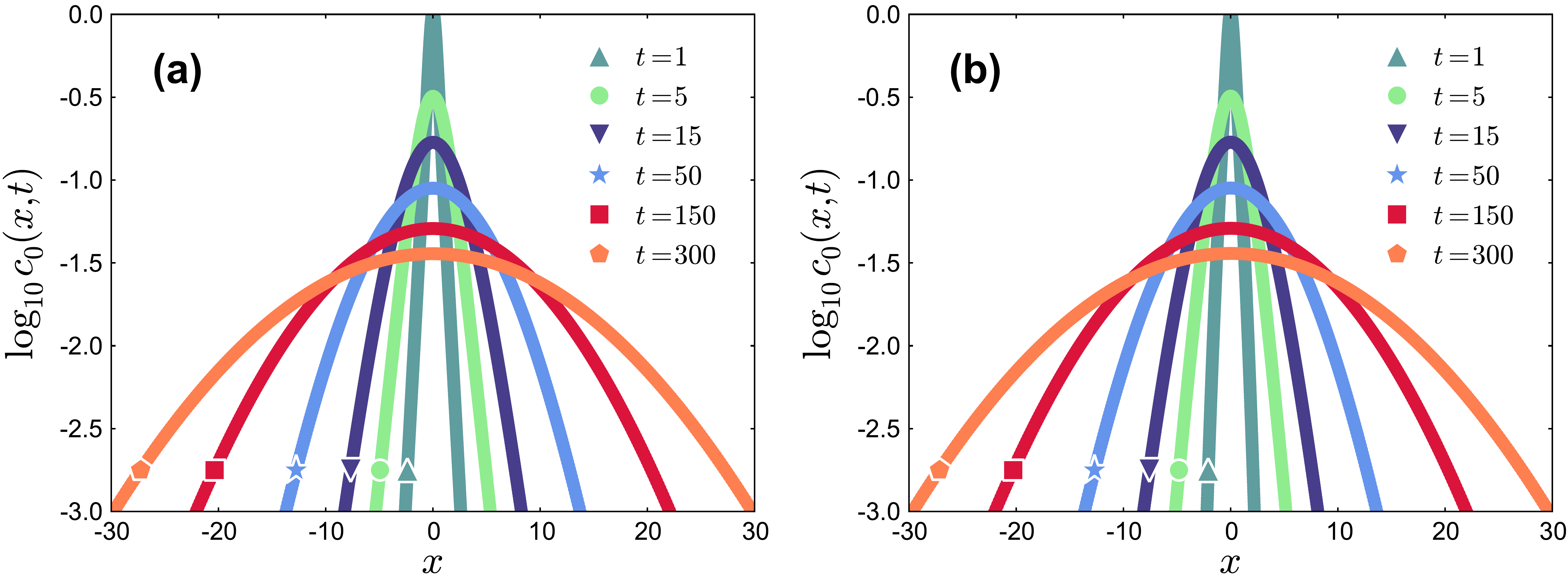}
\caption{\label{fig_directnumeric}(Color online) Semilogarithmic plot of spatial distribution functions $c_0(x,t)$, calculated directly from numerically integrating the Fokker-Planck equation (\ref{FP}) forward in time. For short times exponential tails are observed due to Coulombic friction. They move outward with increasing time, leaving a Gaussian shape in the observation interval. The parameters were set to (a) $\Delta=6$, $K\approx 5.40$ and (b) $\Delta=1.1$, $K=1$. (Technical details: variance of the initial Gaussian spatial distribution $\sigma^2=0.005$; 20000 lattice points in $x$-direction of distance $dx=0.01$; 400 lattice points in $v$-direction of distance $dv=0.05$; $1\times10^4$, $5\times10^4$, $15\times10^4$, $50\times10^4$, $150\times10^4$, $300\times10^4$ time steps of step size $dt=0.0001$, respectively.)}
\end{figure}

In the next step, we numerically integrated eq.~(\ref{FP}) forward in time over the $x$-$v$-space. We started from an initial distribution function $f(x,v,t=0)=\text{``}\delta(x)\text{''}f_{st}(v)$, where ``$\delta(x)$'' was represented by a narrow Gaussian spatial distribution and $f_{st}(v)$ by eq.~(\ref{fst}). The spatial distribution function $c_0(x,t)$ then follows via eq.~(\ref{c0}) at each time step. Again we considered two different scenarios of Coulombic frictional strength of $\Delta=6$, $K\approx5.40$ and $\Delta=1.1$, $K=1$.
Some resulting spatial distribution functions at different times are shown in fig.~\ref{fig_directnumeric}.

For smaller times, exponential tails are identified within the observation window. With increasing time and broadening of the distribution, these exponential tails move outward and the curves within the observation window take on a Gaussian shape. In that sense, we observe a crossover from an exponential to Gaussian behavior.

\begin{figure}
\includegraphics[width=17.cm]{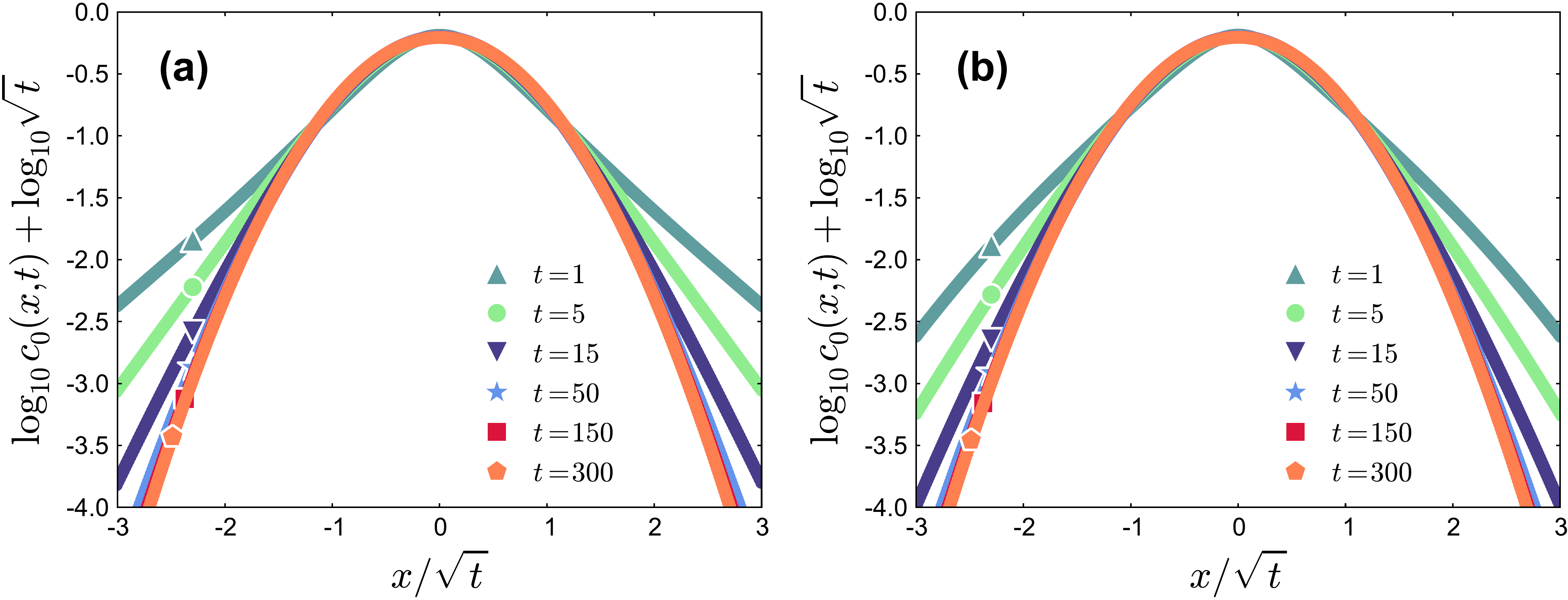}
\caption{\label{fig_datacollapse}(Color online) Simple rescaling of the data curves of fig.~\ref{fig_directnumeric} as given by the axes labels and the main text. The data collapse works well for the Gaussian parts. However, it fails in the exponential regimes. Values of the parameters were set to (a) $\Delta=6$, $K\approx 5.40$ and (b) $\Delta=1.1$, $K=1$.}
\end{figure}

There is a subtle difference in the statistic properties of the transient exponential compared to the final Gaussian behavior. This becomes clear when we try to rescale the different distribution curves in fig.~\ref{fig_directnumeric} to make them collapse onto a single curve. For that purpose, the positions of the values of the distribution function are shifted from $x$ to $x/\sqrt{t}$ for each curve, respectively. The magnitude of the respective distribution function is increased by $\log_{10}\sqrt{t}$ to keep the normalization. As fig.~\ref{fig_datacollapse} shows, a reasonable data collapse is easily achieved for times $t>50$. In that regime, the curves in the observation window are of predominantly Gaussian shape.
To achieve a better data collapse in the central region and stress the presence of the exponential tails for the smaller times, we have used $t=0.4$ instead of $t=1$, $t=3.9$ instead of $t=5$, and $t=14$ instead of $t=15$ to do the rescaling.
However, for the smaller times, the simple data collapse fails.

The reason for this failure becomes evident when we look at the time dependence of the moments of the spatial distribution functions. Since they are even functions in $x$, all uneven moments vanish: $\langle x^n\rangle=0$, if $n$ is an uneven integer.

For the positive even integer values $n$ we write
\begin{equation}
  \langle x^n\rangle\propto t^{\zeta(n)}
\end{equation}
in order to parameterize the breakdown of the simple data collapse. In the Gaussian case, $\zeta(n)/n$ is constant, which we can refer to as singlescaling. This is what we observe for the later times. However, we find that in the early time regime $\zeta(n)/n$ is a function of $n$. In other words, we have found a transient regime that can be represented as multiscaling.

\begin{figure}
\includegraphics[width=17.cm]{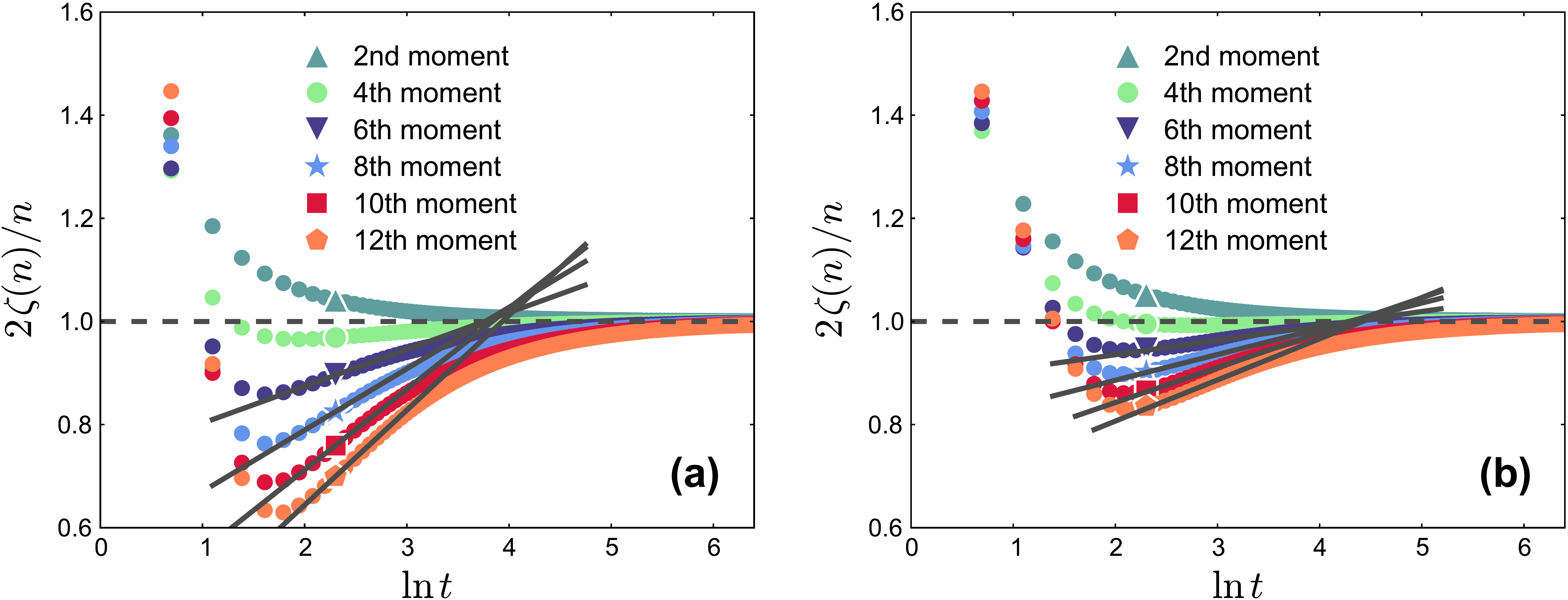}
\caption{\label{fig_zeta}(Color online) $2\zeta(n)/n$ as a function of $\ln{t}$, where the exponent function $\zeta(n)$ is defined by $\langle x^n\rangle\propto t^{\zeta(n)}$. At long times, all curves converge to one value, characteristic for Gaussian statistics. Below a crossover time, however, Coulombic friction leads to explicit non-Gaussian behavior. The two cases correspond to (a) $\Delta=6$, $K\approx 5.40$ and (b) $\Delta=1.1$, $K=1$.}
\end{figure}

Looking at the time dependence of $\zeta(n)$, we can find a crossover time from the exponential to the Gaussian regime. For that purpose, we calculated for each time step the function $\zeta(n)=d(\log{\langle x^n\rangle})/d(\log{t})$, $n=2,4,6,8,10,12$. The resulting time dependent functions $2\zeta(n)/n$ are plotted in fig.~\ref{fig_zeta} for the different values of $n$.

In the Gaussian regime, the values of $2\zeta(n)/n$ should fall on one point at each time step. This becomes true asymptotically at long times $t$, reflecting the data collapse in fig.~\ref{fig_datacollapse} for the long-time distribution functions. We observe that the values of $2\zeta(n)/n$ split for different values of $n$ at small times $t$. This defines an intermediate regime of multiscaling. It results from the influence of the Coulombic frictional term. In fig.~\ref{fig_datacollapse} it corresponds to the non-collapsing curves at smaller times.

\begin{figure}
\includegraphics[width=17.cm]{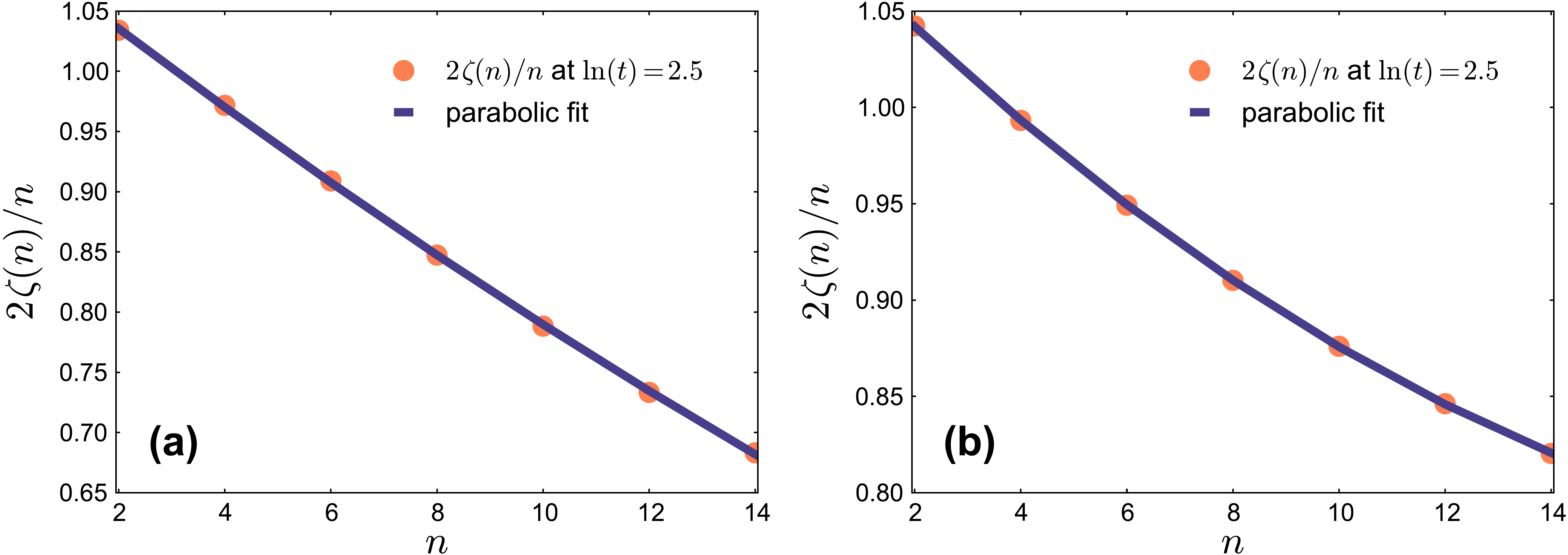}
\caption{\label{fig_zetaline}(Color online) Values of $2\zeta(n)/n$ as a function of $n$ at time $\ln{t}\approx 2.5$ and fitted by a parabola for (a) $\Delta=6$, $K\approx 5.40$ and (b) $\Delta=1.1$, $K=1$.}
\end{figure}

As shown by fig.~\ref{fig_zeta}, in the intermediate regime the values of $2\zeta(n)/n$ relax toward their common Gaussian value. We have fitted their intermediate behavior by straight lines. At $\ln{t}\approx 4$ and $\ln{t}\approx 4.5$, respectively, where these lines approximately cross the asymptotic Gaussian value of $1$, we can roughly define a crossover point between the two regimes. The values of $2\zeta(n)/n$, plotted as a function of $n$, are well fitted by a parabolic curve, as depicted in fig.~\ref{fig_zetaline}.

The initial behavior in fig.~\ref{fig_zeta} for $t\rightarrow0$ is related to the fact that we use a narrow Gaussian distribution as an initial condition. We have checked our numerics by comparing to the case where Coulombic friction is absent ($\Delta=0$, $K=1$). Then Gaussian behavior is obtained at all times: all values of $2\zeta(n)/n$ for different $n$ fall onto one data point at each time step $t$.

In conclusion, the nonlinear Coulombic frictional force leads to apparent multiscaling on intermediate time scales. This is revealed by the failure of a simple data collapse when the data are rescaled only on a single scale. It becomes even more evident from the varying time dependence of the different moments of the spatial distribution function.

\section{\label{discussion}Discussion}

In this section, we shortly discuss the observability of Coulombic frictional effects during experiments on diffusive motions. More precisely, we refer to particle tracking in systems at thermal equilibrium. 
The usual quantity evaluated in such experiments is $\langle x^2\rangle$, the mean square displacement. A diffusion coefficient $D$ is then derived from the linear increase of the mean square displacement with time, $\langle x^2\rangle=2Dt$. 

\begin{figure}
\includegraphics[width=17.cm]{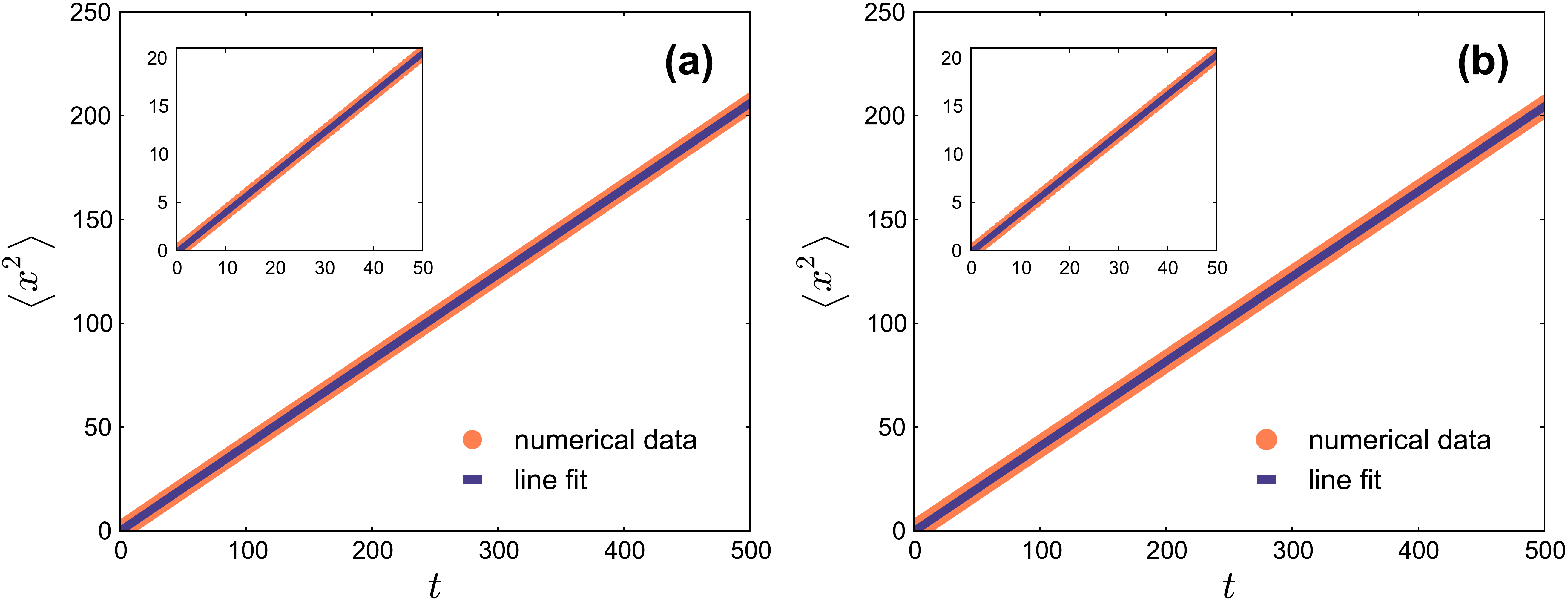}
\caption{\label{fig_msd}(Color online) Mean square displacement as a
function of time for (a) $\Delta=6$, $K\approx 5.40$ and (b)
$\Delta=1.1$, $K=1$. We find a linear relationship $\langle
x^2\rangle\propto t$ in the major part of the inspected time interval.
The inset (same axes labels) stresses that this linear relationship
also prevails at the early times where we observed the exponential
tails within the observation window.}
\end{figure}

We have calculated the time dependence of the mean square displacement as the second moment of the spatial displacement distribution functions. The results are shown in fig.~\ref{fig_msd} for the same parameter values as for those used to obtain figs.~\ref{fig_c0}-\ref{fig_zetaline}. 
As in the case of conventional Brownian motion, we find a linear increase of the mean square displacement during the whole time of observation. 

Consequently, to clarify
the nature of the underlying frictional mechanism during an experiment, the evaluation of the mean square displacement alone is not sufficient. It would be important to evaluate the higher order moments from the
measured distribution functions. Although the $12th$ moment listed
in fig.~\ref{fig_zeta} is certainly not realistic, the fourth or maybe
sixth moment should be possible. A spreading of the function
$2\zeta(n)/n$ for different values of $n$ would reveal more about the
underlying frictional process.

In this context, we mention two experiments on soft matter systems. In both cases, a linear increase of the mean square displacement was found. An analysis of the higher order moments as outlined above may be interesting. 

The first example is the diffusive motion of
tethered vesicles on supported lipid bilayers
\cite{yoshina2003arrays,yoshina2006diffusive}. The vesicles were
tethered by DNA-strands chemically attached to the hydrophilic head
groups of single lipid molecules. One of these lipids was part of the
vesicle membrane, the other was part of the lipid bilayer.
Interestingly, the properties of the motion could be changed by
controlled addition of salts or polymers.

It was pointed out that effective frictional
forces may play a significant role for the motion observed. These
frictional forces should mainly result from the interaction between the
vesicles and the bilayer membrane (and probably not from
dragging the anchoring lipid molecule through the supported bilayer
membrane) \cite{yoshina2006diffusive}. Besides, intermediate trapping
by defects in the supporting bilayer should slow down the diffusive
motion of the vesicles. An analysis of the higher order moments and the
corresponding exponents $\zeta(n)$ could reveal more about the
contributions of each of these processes.

The other example is the one-dimensional diffusive motion of colloidal
polystyrene particles on top of straight bilayer membrane tubes \cite{wang2009anomalous}. These tubes were adsorbed on a solid substrate and not moving in the lateral direction. On top of the tubes, particles of roughly the same
diameter ($\sim 100$ nm) were adsorbed. They were repelled from the
substrate due to electrostatic interactions. Their motion in a fluid
environment (mainly water) was recorded by methods of single particle
tracking \cite{anthony2006methods}. The time dependent spatial distribution
functions were found to have the same qualitative appearance in shape as the ones
shown in fig.~\ref{fig_directnumeric}. In particular, a crossover from
exponential to Gaussian shape of the distribution function was
identified within the observation window. The adsorption on top of the
tubes reduced the diffusion coefficient $D$ by a factor of $1/5$ when compared to the non-adsorbed case. 

We must remark at this point that the length and time scales accessible by our numerical calculations differ by several orders of magnitude from the ones in these experiments. In particular, it is not clear by which speed the exponential tails of the distribution function would move outward on these experimental time scales. (In other words, on the experimental time scales the exponential tails due to an effective Coulombic friction may have left the observation window already, and the exponential tails observed in ref.~\cite{wang2009anomalous} may have a different source). At this stage, a direct connection between our results and the above experimental results is therefore not possible. Again, this stresses the importance of analyzing the higher order moments of the experimentally obtained distribution functions. 

The sets of parameter values (a) $\Delta=6$, $K\approx 5.40$ and (b) $\Delta=1.1$, $K=1$ used in figs.~\ref{fig_c0}-\ref{fig_msd} both lead to a reduction of the diffusion coefficient $D$ by a factor of $1/5$ when compared to the absence of Coulombic friction $\Delta=0$, $K=1$. On the one hand, in case (a), we followed the fluctuation dissipation relation eq.~(\ref{flucdiss}). On
the other hand, in case (b), we kept the strength of the stochastic force $K$ the
same as without Coulombic friction. These are the two scenarios outlined at the end of section \ref{equations}. The crossover value $\Delta^*$ is calculated for the two scenarios as $\Delta^* \approx 2.33$ and $\Delta^*=1$, respectively. Since $\Delta>\Delta^*$, both cases probe the partly stuck regime (see fig.~\ref{fig_phasediagram}). 

We want to close this discussion with general remarks on the nature of
eq.~(\ref{langevinv}). As always when dealing with frictional behavior
in a phenomenological way \cite{persson2000sliding},
eq.~(\ref{langevinv}) should be regarded as a simple qualitative
approach to effectively describe the more complicated underlying
processes at a lower level of description. From underlying rate
processes, the correct stochastic differential equation can be
described.

When the Langevin equations (\ref{langevinv}) and (\ref{langevinx}) are
used to characterize experimental results, only two degrees of freedom,
$x$ and $v$, are retained from the underlying many-body system
description. In general, however, they are coupled to the other degrees
of freedom of the system. Therefore, we implicitly assume that these
other degrees of freedom have been integrated out (see
refs.~\cite{feynman1963theory,caldeira1981influence,caldeira1983path,caldeira1983quantum}
for a well-known treatment). The coupling to the environment is only
reflected in the values of the parameters $\tau$, $\Delta$, and $K$, as
well as the functional form of the frictional and stochastic forces. We
must perform this integration procedure explicitly, if we wish to
obtain detailed functional forms for the frictional and stochastic
forces from more microscopic models. Such a procedure would determine
the precise relation between the strengths of the frictional and
stochastic forces.

Here, we have followed a purely phenomenological approach. We discussed two cases for the strength of the stochastic force. On the one hand,
the fluctuation-dissipation relation eq.~(\ref{flucdiss}) marks an upper limit for the strength of the stochastic force $K$ in a system at thermal equilibrium (see the appendix). On the other hand, the value $K=1$ corresponds to its value in the absence of Coulombic friction. Within the narrow window accessible to the direct numerical calculations, both of these two scenarios lead to qualitatively identical results.

\section{\label{conclusion}Conclusions}

The effect of Coulombic frictional forces in stochastic equations of
motion formed the central topic of this paper. We have mainly studied
the statistical properties of the spatial distribution functions
resulting from the related Fokker-Planck equation.

As a first step, we have revealed the connection of the underlying
equation to the case of a quantum mechanical oscillator in the presence
of a pinning delta-potential. This allowed us to derive the
coefficients in the corresponding Brinkman hierarchy. The latter then
offers the possibility of numerically calculating the spatial
distribution functions in an approximate but time-efficient way.
In addition, we numerically integrated the Fokker-Planck equation forward
in time directly.

We have found exponential tails in the spatial distribution functions
that move out of the observation window with increasing time. In the
intermediate time regime, where these exponential tails are present,
effective multiscaling has been detected for the spatial distribution functions.
The data collapse through a simple rescaling procedure fails. This
becomes more evident when the time dependence of the different moments
of the distribution is inspected and contrasted with the purely
Gaussian case. Such a procedure  serves to identify the regime where
Coulombic friction dominates, and a crossover time can be extracted.

Finally, we have discussed the possibility of observing the influence of Coulombic friction during experiments on diffusive motions. Our central
conclusion is that it is important to determine and analyze the
behavior of the higher order moments of the experimental distribution
functions to learn more about the underlying physical processes.

After this work had been completed, a related manuscript appeared
\cite{touchette2010brownian}. The authors of this manuscript also find
the eigenfunctions to eq.~(\ref{langevinv}) and present details on the
analysis of the corresponding spectra. A spatial component
corresponding to eq.~(\ref{langevinx}), however, is not taken into
account in ref.~\cite{touchette2010brownian}.

\begin{acknowledgments}
We thank Bo Wang and Steve Granick for stimulating discussions on their
experiments at the beginning of this study. Support of this work by the
Deutsche Forschungsgemeinschaft through a research fellowship (A.M.M.)
is gratefully acknowledged.
\end{acknowledgments}

\appendix*

\section{Fluctuation dissipation relation}

In this appendix, we discuss the issue of deriving a fluctuation dissipation relation for a system in thermal equilibrium that obeys the Langevin equations (\ref{langevinv}) and (\ref{langevinx}), or the corresponding Fokker-Planck equation (\ref{FP}). In other words, we are looking for an expression for the strength of the stochastic force $K$ as a function of the friction parameters (we restrict ourselves to constant values of $K$ in this study). For that purpose it is sufficient to concentrate on the velocity dependent part of the equations. Then the only degree of freedom in the model is the velocity of the particle $v$. 

We start from the equipartition theorem 
\begin{equation}
\left\langle v\frac{\partial H}{\partial v} \right\rangle = 1
\end{equation}
in rescaled units. $H$ is the Hamiltonian that characterizes the current energetic state of the system. 

Performing the ensemble average using the stationary velocity distribution function $f_{st}(v)$ from eq.~(\ref{fst}), we obtain the condition 
\begin{equation}
\int_{-\infty}^{\infty}dv \,e^{-\frac{1}{K}\left(\frac{v^2}{2}+\Delta|v|\right)} 
  v\frac{\partial H}{\partial v} 
  =  \sqrt{2\pi K}
      e^{\frac{\Delta^2}{2K}}\left(1-\text{erf}\left\{\frac{\Delta}{\sqrt{2K}}\right\}\right).
\end{equation}
This relation is satisfied, if the Hamiltonian reads
\begin{equation}
H(v) = \frac{1}{K}\left(\frac{v^2}{2}+\Delta|v|\right).
\label{H}
\end{equation}

At this point, we must note that $H(v)$ does not correspond to the kinetic energy of the particle itself, although $v$ describes its velocity. Eqs.~(\ref{langevinv})-(\ref{FP}) and (\ref{H}) must rather be interpreted as effective phenomenological equations for one single remaining degree of freedom $v$. ``Effective'' here means that all the other degrees of freedom that are present in more microscopic characterizations and describe the energetic state of the environment of the particle have been integrated out. The energetic effect on the environment due to frictional interactions, however, is expressed as a function of the magnitude of $v$ and enters into expression (\ref{H}). 

In the viscous case of a freely moving Brownian particle ($\Delta=0$), the Hamiltonian $H(v)$ is set equal to the kinetic energy of the particle. From eq.~(\ref{H}) we then find the usual textbook example $K=1$ (or $K=1/\tau$, if we do not scale out the viscous relaxation time $\tau$) \cite{zwanzig2001nonequilibrium}. Here, since deriving microscopic models is beyond the scope of this study, we do not know the value of the proportionality constant $K$. What we do know on physical grounds is that the mean kinetic energy of the particle itself, $\langle v^2/2\rangle$, in a passive system cannot be larger in the presence of Coulombic friction than without it:
\begin{equation}
\left\langle{v^2}\right\rangle \leq 1. 
\end{equation}
Performing the ensemble average, again using $f_{st}(v)$ from eq.~(\ref{fst}), leads to the fluctuation dissipation relation (\ref{flucdiss}) and thus sets an upper limit for the strength of the driving stochastic force $K$.

\end{document}